\documentclass[aps,prl,twocolumn,showpacs,superscriptaddress]{revtex4-1}  

\usepackage{graphicx}  
\usepackage{dcolumn}   
\usepackage{bm}        
\usepackage{amssymb}   
\usepackage{mathtools}
\usepackage{etoolbox}

\appto\normalsize{\belowdisplayshortskip=\belowdisplayskip}
\appto\small{\belowdisplayshortskip=\belowdisplayskip}
\appto\footnotesize{\belowdisplayshortskip=\belowdisplayskip}
\usepackage[utf8]{inputenc}
\DeclareMathSizes{5}{13}{7}{3}

\begin{document}

\title{The phonon Hall viscosity of ionic crystals}

\begin{abstract}
When time-reversal symmetry is broken, the low-energy description of acoustic lattice dynamics allows for a dissipationless component of the viscosity tensor, the phonon Hall viscosity, which captures how phonon chirality grows with the wavevector. In this work, we show that, in ionic crystals, a phonon Hall viscosity contribution is produced by the Lorentz forces on moving ions. We calculate typical values of the Lorentz force contribution to the Hall viscosity using a simple square lattice toy model, and we compare it with literature estimates of the strengths of other Hall-viscosity mechanisms. 
\end{abstract}
\author{B. Flebus}
\affiliation{Department of Physics, Boston College, 140 Commonwealth Avenue Chestnut Hill, MA 02467}
\author{A.H. MacDonald}
\affiliation{Physics Department, University of Texas at Austin, Austin TX 78712}

\date{\today}
\maketitle

\textit{Introduction.}
Recent measurements of giant thermal Hall signals in many insulating ionic crystals~[\onlinecite{exp1,exp2,exp3,exp4,exp5}] have ignited widespread interest in the processes underlying chiral phonon transport. The mechanisms by which 
phonons acquire chirality can be broadly divided in two classes: 
i) intrinsic - {\it i.e.}, originating from external magnetic fields or 
magnetism that breaks time-reversal symmetry in crystals ~[\onlinecite{1PhononHall, sachdev1,1PhononHall2,chiralspinons,electron,1PhononHall,electron1,electron2,electron3,electron4,electron5,spin1,spin2,spin3,spin4,spin5Berry, spin6Berry, spin7Berry,magnon1,magnon2,spinon}], and ii) extrinsic - {\it i.e.}, originating from scattering on time-reversal-symmetry-breaking crystal defects~[\onlinecite{extrinsic,defect2,defect3}]. 

In the low-energy elasticity-theory description of acoustic waves, 
intrinsic time-reversal symmetry breaking is 
accounted for by a Hall viscosity contribution
to the response of the system’s viscoelastic stress tensor to an applied strain $u_{ij}$~\cite{1PhononHall,1PhononHall2}:
\begin{align}
\langle \hat{T}_{ij} \rangle = \Lambda_{ijkl} \, u_{kl} + \eta_{ijkl} \, \dot{u}_{kl}\,.
\end{align}
Here $\hat{T}$ is the stress tensor, $u_{ij}=\frac{1}{2}\left( \partial_{i} 
u_{j} + \partial_{j} u_{i} \right)$ is the 
strain tensor, $u_{i}$ is the atomic displacement field, and $\Lambda$ and $\eta$ are, 
respectively, the elasticity and viscosity tensors.  
The viscosity tensor $\eta$ can have dissipationless component, dubbed the Hall (or odd) viscosity, which is associated with the part of $\eta_{ijkl}$ that is antisymmetric  under exchange of the pairs of indices $(ij)$ and $(kl)$.

Dissipationless Hall contributions to the 
viscosity [\onlinecite{1PhononHall}]  are allowed only in systems with broken time-reversal symmetry 
where they alters the acoustic phonon spectrum, 
and mix longitudinal and transverse modes.
\begin{figure}[b!]
\includegraphics[width=0.75\linewidth]{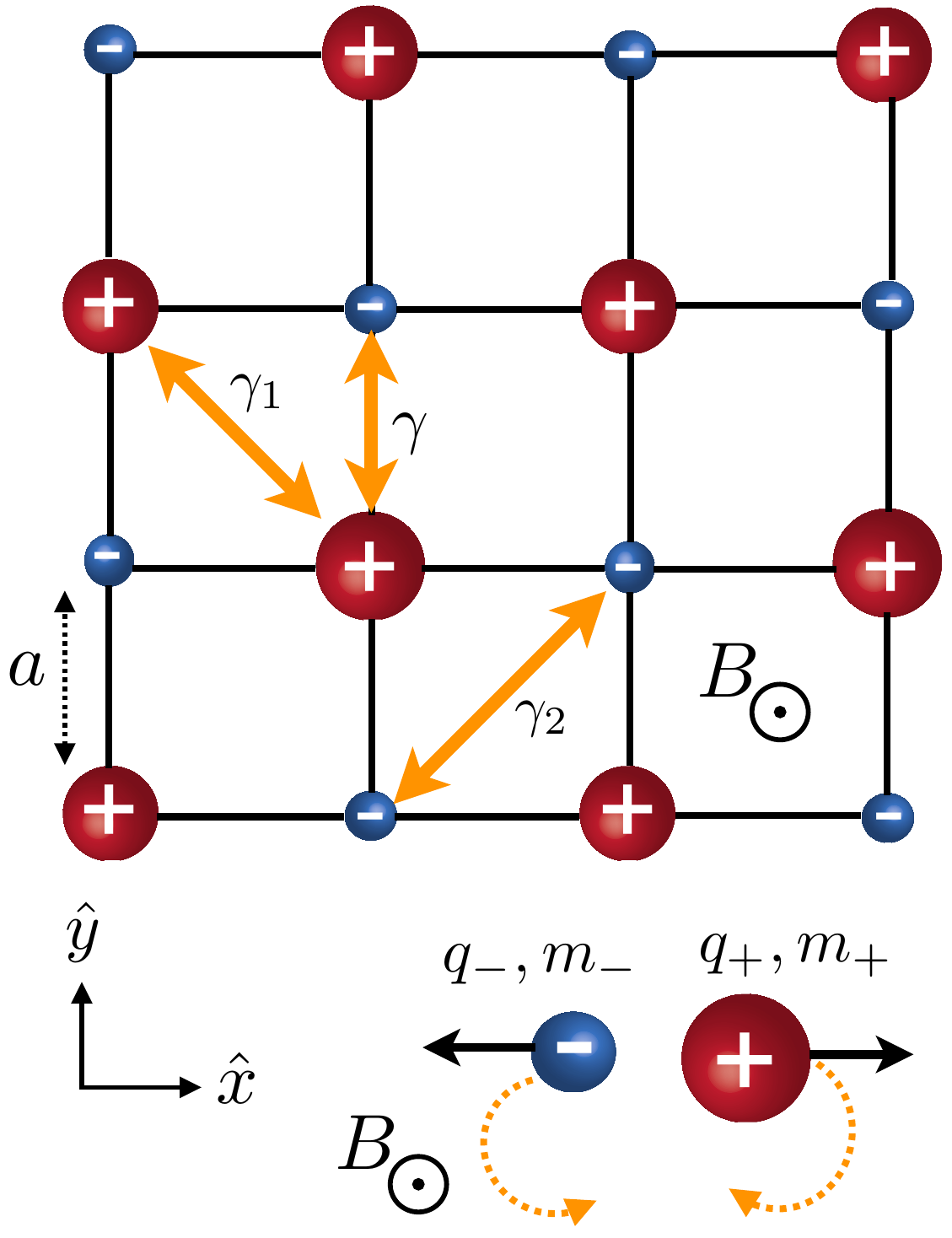}
\caption{ Top: Diatomic square lattice with interatomic distance $a$, subjected to an out-of-plane magnetic field $B$. The cation
and anion masses are $m_{+}$ and $m_{-}$, and the ion effective charges are $q_{\pm}=\pm Z^*e$.  
We include centro-symmetric pairwise interactions between ions characterized by spring constants 
$\gamma$, $\gamma_{1}$ and $\gamma_{2}$ for near-neighbor interactions between
cations and  anions, second neighbor interactions cations, and second neighbor interactions between
anions, respectively. Bottom:  Optical modes in an ionic crystal. In the presence of a magnetic field $B$ perpendicular to the $2d$ crystalline plane,   the Lorentz forces on cations and anions moving in opposite directions do not cancel.}
\label{Fig1}
\end{figure} 
Previous theoretical work has addressed phonon Hall viscosities produced by
 coupling of acoustic phonons to a ferroelectric~[\onlinecite{electron}], electronic~[\onlinecite{ sachdev1,1PhononHall,chiralspinons,electron1,electron2,electron3,electron4,electron5}] 
or spin environment~[\onlinecite{spin1,spin2,spin3,spin4,spin5Berry, spin6Berry, spin7Berry,magnon1,magnon2,spinon}]
in which time-reversal symmetry is broken.

In this Letter, we consider the role of
Lorentz forces in ionic crystals, which act on the electric dipoles produced 
by out-of-phase motion of cations and anions. The out-of-phase lattice vibrations are, in turn, coupled at finite wavevector to the in-phase modes by elastic forces.  By deriving an effective theory for the low-energy acoustic waves, 
 we show that Lorentz forces always lead to finite Hall viscosities that are linear in 
magnetic field.  

We employ a two-dimensional ($2d$)
toy model, \textit{i.e.}, the diatomic square lattice subjected to an out-of-plane magnetic field sketched in Fig.~\ref{Fig1},
to estimate the typical size of contributions to the Hall viscosity tensor $\eta$
supplied by this mechanism.  
Due to the $C_4$ rotational symmetry of our $2d$ model, there is only one independent coefficient in the phonon Hall viscosity tensor, \textit{i.e.},  $\eta^{H}=\eta_{xxxy}$~[\onlinecite{Landau}].  
We find that this coefficient is proportional to the external magnetic field,
and that its typical numerical value is comparable to the those 
estimated for other mechanisms.  Our results show that every ionic crystal can display a phonon Hall response, independently of phonon coupling to external degrees of freedom. 
\\ \\ \indent
\textit{Model.}
We consider a square lattice with cations and anions on opposite sublattices that is 
subjected to an out-of-plane magnetic field $B$, as depicted in Fig.~\ref{Fig1}. 
The cation and anion have masses $m_{+}$ and $m_{-}$ and charges $q_{\pm}=\pm Z^* e$, respectively, 
with $e>0$ being the electron charge and $Z^*$ the effective ionic charge number. 
We assume centro-symmetric forces with spring constants $\gamma$, $\gamma_{1}$ and $\gamma_{2}$ 
for interactions between cation and anion nearest-neighbors, cation second nearest-neighbors,
and anion second nearest-neighbors, respectively. 

At wavevector $\mathbf{k}$, the lattice displacement along the $i$th direction
of the $n$th ion (with $n=\pm$) in the $l$th cell can be written as
\begin{align}
u_{inl}(\mathbf{k},\omega)=m^{-1/2}_{n} u_{in} e^{  i  \mathbf{k} \cdot \mathbf{R}_{n}(l)-i \omega t}\,.
\end{align}
 Here, $u_{in}$ is the lattice vibration amplitude of the $n$th atom along the $i$th direction, $\mathbf{R}_{n}(l)$ is a lattice translation vector, and
 $\omega$  the normal mode  frequency.
In the harmonic approximation, the equation of motion for the vibration amplitude $u_{in}$ is 
\begin{align}
\omega^{2} u_{ni}=\sum_{m,j} D_{ij}(mn,\mathbf{k}) u_{mj} + \sum_{j} \frac{i\omega B q_{n}}{m_{n}}  u_{nj} \epsilon_{ij}\,,
\label{eqmotion}
\end{align}
where $\epsilon_{ij}$ is the 2$d$ Levi-Civita tensor ($\epsilon_{\pm \mp}=\pm 1$ and $\epsilon_{\pm \pm}=0$). 
The element $D_{ij}(mn,\mathbf{k})$ of the dynamical matrix reads as
\begin{align}
D_{ij}(mn,\mathbf{k})=-\sum_{l'} \frac{ \gamma_{nm}}{\sqrt{m_{m} m_{n}}} e_{i}(m)e_{j}(m)  e^{i \mathbf{k} \cdot \mathbf{R}_{m}(l')}\,,
\label{dynamicalmatrix}
\end{align}
where $\hat{e}(m)$ is the unit vector along $\mathbf{R}_m(l')$
and $\gamma_{nm}$ is the spring constant between $n$th and $m$th ions. 
Plugging Eq.~(\ref{dynamicalmatrix}) into Eq.~(\ref{eqmotion}), we can rewrite the equations of motion as
\begin{align}
\omega^2  \begin{pmatrix}
\mathbf{u}_{+} \\ \mathbf{u}_{-}
\end{pmatrix}= \mathcal{A}(\mathbf{k},\omega) \begin{pmatrix}
\mathbf{u}_{+} \\ \mathbf{u}_{-}
\end{pmatrix}\,,
\label{eqmotion}
\end{align}
where 
\begin{widetext}
\begin{align}
\mathcal{A}(\mathbf{k},\omega)=\begin{bmatrix} \frac{2\gamma + 2\gamma_{1}\left(1 -\cos k_{x}a \cos k_{y}a  \right)}{m_{+}} &  \frac{i\omega Ze B-2\gamma_{1}\sin k_{x}a \sin k_{y}a}{m_{+}} &-\frac{2\gamma  \cos k_{x}a}{\sqrt{m_{+} m_{-}}} & 0  \\ -\frac{2\gamma_{1}\sin k_{x}a \sin k_{y}a + i\omega Ze B}{m_{+}} & \frac{2\gamma+ 2\gamma_{1}\left(1 -\cos k_{x}a \cos k_{y}a  \right)}{m_{+}}  & 0 &   -\frac{2\gamma \cos k_{y}a }{\sqrt{m_{+} m_{-}}} \\ -\frac{2\gamma \cos k_{x}a }{\sqrt{m_{+} m_{-}}}  & 0 & \frac{2\gamma+ 2\gamma_{2} \left(1 -\cos k_{x}a \cos k_{y}a  \right)}{m_{-}}  & -\frac{2\gamma_{2} \sin k_{x}a \sin k_{y}a+i\omega Ze B}{m_{-}}   \\ 0 & -\frac{2\gamma \cos k_{y}a}{\sqrt{m_{+} m_{-}}}   &   \frac{i\omega Ze B-2\gamma_{2} \sin k_{x}a \sin k_{y}a}{m_{-}} &\frac{2\gamma+2\gamma_{2}\left(1 -\cos k_{x}a \cos k_{y}a  \right)}{m_{-}}   \end{bmatrix} \,,
\label{96}
\end{align}
\end{widetext}
with $a$ being the  lattice constant. 
 In the following, we set $k_{y}=0$ and we focus on the long-wavelength limit of the phonon dynamics by
 assuming that $k_{x}a\ll1$. It is convenient to work in the basis of eigenmodes of Eq.~(\ref{eqmotion}) for $k_{x}=0$ and $B=0$, which are the in-phase ($\mathbf{u}_{a}$) and out of phase ($\mathbf{u}_{o}$) ionic motions,
\begin{align}
u_{ai}=&\sqrt{\frac{m_{-}}{m_{-}+m_{+}}} \left( u_{-i} + \sqrt{\frac{m_{+}}{m_{-}}} u_{+i} \right)\,,
\label{102}
\end{align}
\begin{align}
u_{oi}=&\sqrt{\frac{m_{+}}{m_{-}+m_{+}}} \left( u_{-i}- \sqrt{\frac{m_{-}}{m_{+}}} u_{+i} \right)\,.
\label{103}
\end{align}

In this basis, we can rewrite the dynamical matrix~(\ref{96})  as

\begin{widetext}
\begin{align}
\resizebox{.99\hsize}{!}{$
\mathcal{A}(k_{x},\omega)=\begin{bmatrix} \frac{\left( 2\gamma + \gamma_{1}+\gamma_{2}\right) (k_{x}a)^2}{m_{+}+m_{-}} & 0 & \frac{\left[ m_{+} (\gamma+\gamma_{2}) - m_{-} (\gamma+\gamma_{1}) \right] (k_{x}a)^2}{\sqrt{m_{+}m_{-}} (m_{+}+m_{-})} & -\frac{iZe B\omega}{\sqrt{m_{+} m_{-}}} \\ 0 & \frac{\left(  \gamma_{1}+\gamma_{2}\right) (k_{x}a)^2}{m_{+}+m_{-}} & \frac{iZe B\omega}{\sqrt{m_{+} m_{-}}} & \frac{\left( m_{+} \gamma_{2}-m_{-} \gamma_{1}\right) (k_{x}a)^2}{\sqrt{m_{+} m_{-}}(m_{+}+m_{-})}  \\ \frac{\left[ m_{+} (\gamma+\gamma_{2}) - m_{-} (\gamma+\gamma_{1}) \right] (k_{x}a)^2}{\sqrt{m_{+}m_{-}}(m_{+}+m_{-})} & -\frac{iZe B\omega}{\sqrt{m_{+} m_{-}}} & \frac{2\gamma(m_{+}+m_{-})^2 + ( \gamma_{2} m^2_{+} + \gamma_{1} m^2_{-} -2\gamma m_{+} m_{-})(k_{x}a)^2}{m_{+} m_{-} (m_{+}+m_{-})}&   - \frac{i Ze B (m_{+}-m_{-})\omega}{m_{+}m_{-}} \\ \frac{iZe B\omega}{\sqrt{m_{+} m_{-}}} & \frac{\left( m_{+} \gamma_{2}-m_{-} \gamma_{1}\right) (k_{x}a)^2}{\sqrt{m_{+} m_{-}}(m_{+}+m_{-})}    & \frac{i Ze B (m_{+}-m_{-})\omega}{m_{+}m_{-}} & \frac{2\gamma(m_{+}+m_{-})^2 + ( m^2_{+} \gamma_{2}+ m^2_{-} \gamma_{1}) (k_{x}a)^2}{m_{+} m_{-} (m_{+}+m_{-})}\end{bmatrix}$}\,.
\label{96newbasis}
\end{align}
\end{widetext}

Equation~(\ref{96newbasis}) shows that the in-phase motion is coupled to the out-of-phase motion  when
either $k_{x}$ or $B$ is non-zero. The longitudinal (transverse) in-phase motion of the cations and anions is coupled  to out-of-phase longitudinal (transverse) motion via elastic forces, which vanish as $k_{x} \rightarrow 0$.  In contrast the Lorentz force yields a wavevector-independent interaction between the longitudinal in-phase motion and the out-of-phase transverse motion.  When the cation and anion have different masses, the Lorentz force also  directly couples the transverse and longitudinal out-of-phase motions. 

\textit{Phonon Hall viscosity}.
We derive a low-energy theory for the in-phase modes $\textbf{u}_{a}$
by integrating over out--of-phase modes in the imaginary-time phonon-system 
action $\mathcal{S}\left[ \mathbf{u}_{a}, \mathbf{u}_{o} \right]$ corresponding to
Eq.~(\ref{96newbasis}). The effective action $\mathcal{S}_{a}$ for the low-energy nearly 
in-phase modes is 
\begin{align}
e^{-\mathcal{S}_{a}[\mathbf{u}_{a}]}=\int \mathcal{D} \mathbf{u}^{*}_{o} \; \mathcal{D} \mathbf{u}_{o} \; e^{-\mathcal{S}[\mathbf{u}_{a},\mathbf{u}_{o}]}\,.
\end{align}
\begin{figure}
\includegraphics[width=0.9\linewidth]{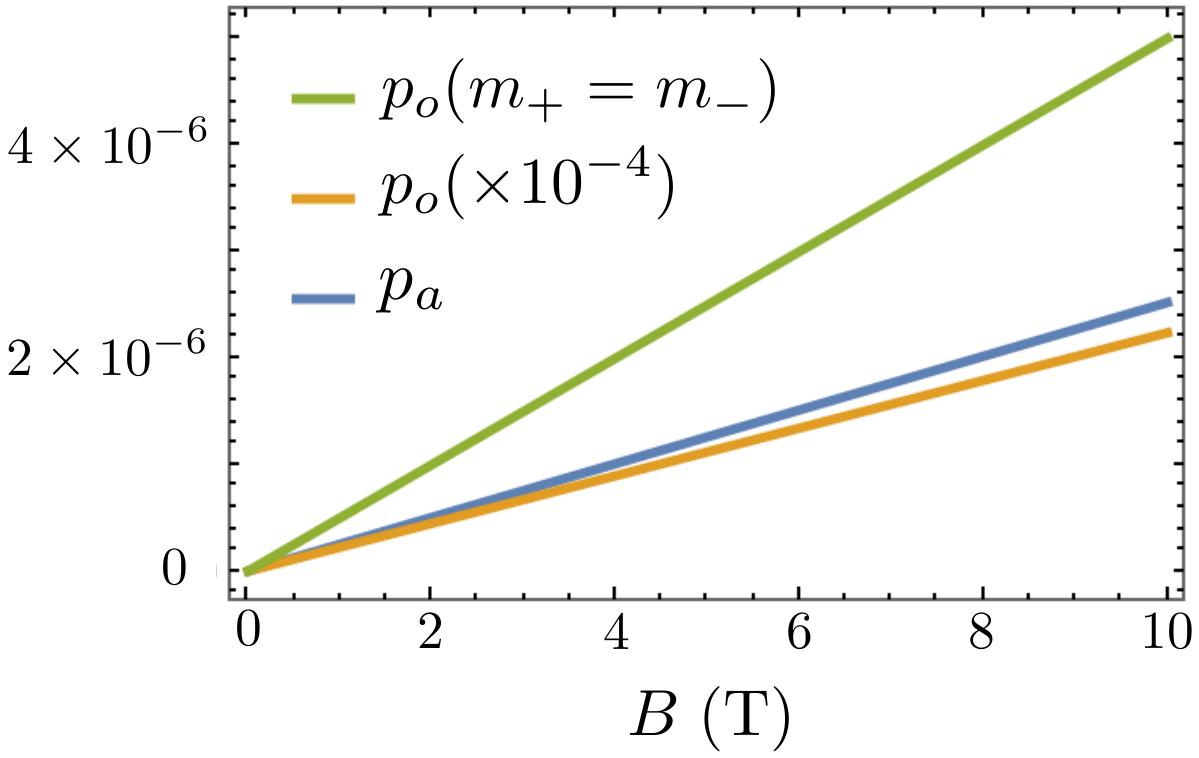}
\caption{Blue line: Dependence of the polarization $p_{a}$ of the longitudinal in-phase mode  on the magnetic field, calculated from Eq.~(\ref{pol}) and  Eq.~(\ref{96newbasis}). The two sets of data overlap. Orange and green line: Dependence of the polarization $p_{o}$ of the longitudinal out-of-phase mode on the magnetic field when choosing Cu and O as cation and anion masses and when setting $m_{+}=m_{-}$, respectively. 
If not otherwise specified, the figures are plotted using  parameters listed in the main text.}\label{Fig:2}
\end{figure}
To leading order in the small parameters $k_{x}a$ and $\omega_{c}/\omega_{o}$, the resulting 
equations of motion are,
 \begin{align}
&\omega^2 \left[ 1 +\frac{\omega^2_{c}}{\omega^2_{o}}\right] u_{ax}= c^2_{l} (k_{x}a)^2 u_{ax} \nonumber \\
& +\frac{i\omega \omega_{c}}{\omega^2_{o}}  \frac{\left[ m_{+} (\gamma+2\gamma_{2}) -m_{-} (\gamma + 2 \gamma_{1})\right] (k_{x}a)^2}{(m_{+}+m_{-}) \sqrt{m_{+} m_{-}}}
 u_{ay}\,, 
 \label{eqmotioneff1}
\end{align} 
 \begin{align}
& \omega^2 \left[ 1 +\frac{\omega^2_{c}}{\omega^2_{o}} \right] u_{ay}= c^2_{t} (k_{x}a)^2 u_{ay}\nonumber \\
 &-\frac{i\omega \omega_{c}}{\omega^2_{o}}  \frac{\left[ m_{+} (\gamma+2\gamma_{2})-m_{-} (\gamma + 2 \gamma_{1})  \right] (k_{x}a)^2}{(m_{+}+m_{-}) \sqrt{m_{+} m_{-}}} u_{ax}\,. 
 \label{eqmotioneff2}
 \end{align}
Here $\omega_{o}$ is the optical phonon frequency 
at the $\Gamma=(0,0)$ point, $\omega_c$ is the cyclotron frequency,
and $c_{l}$ and $c_{t}$ are the longitudinal and transverse phonon velocities:
\begin{align}
&\omega_{o}=\sqrt{\frac{2\gamma (m_{+}+m_{-})}{m_{+} m_{-}}}\,, \; \; \;  \omega_{c}= \frac{Z^{*}e B}{\sqrt{m_{+} m_{-}}}\,, \nonumber \\
&c_{l}=\sqrt{\frac{2\gamma + \gamma_{1}+\gamma_{2}}{m_{+}+m_{-}}}\,, \; \; \; \; \;\; \; \; c_{t}=\sqrt{\frac{\gamma_{1}+\gamma_{2}}{m_{+}+m_{-}}}\,.
\label{Eq10}
\end{align}

The phonon Hall viscosity $\eta_{H}$ can be extracted from Eqs.~(\ref{eqmotioneff1}) and~(\ref{eqmotioneff2}) 
by identifying $\rho=a^{-2}\left( m_{+} + m_{-} \right)$ 
as the mass density of the system and adopting the definition of phonon Hall viscosity introduced in the seminal work of Barkeshli \textit{et al.}~[\onlinecite{1PhononHall}].  We find that 
\begin{align}
\eta_{H}
=&\frac{\omega_{c}}{\omega_0^2}\frac{m_{+}\left( \gamma + 2\gamma_{2} \right)-m_{-}\left(\gamma+2\gamma_{1} \right)}{  \sqrt{m_{-} m_{+}}}\,, \nonumber \\
=& \frac{m_{+}\left( \gamma + 2\gamma_{2} \right)-m_{-}\left(\gamma+2\gamma_{1} \right)}{  2\gamma (m_{+}+m_{-})} Z^{*} e B\,.
\label{Eq265}
\end{align}
Equation~(\ref{Eq265}) shows that acoustic phonons in ionic crystals can acquire a dissipationless Hall viscosity whose strength is proportional to the magnetic field. This is the central result of our work.

The strength and sign of the phonon Hall viscosity~(\ref{Eq265}) strongly depends on the ratio $x=m_{-}/m_{+}$ between the mass of cations and anions. 
For $x\gg 1$, the Hall viscosity is negative and approaches the value $\eta_{H} \sim -\left( 1/2 + \gamma_{1}/\gamma \right) ZeB$. For $x=(\gamma+2\gamma_{2})/(\gamma + 2\gamma_{1})$ $\eta_{H}$ crosses zero, to then increase until it reaches its saturation value $\eta_{H} \sim \left( 1/2 + \gamma_{2}/\gamma \right) ZeB$ for $x\ll 1$.  We therefore expect positive  Hall viscosities in oxides because 
their anions are light.  

Since the Lorentz force couples transverse and longitudinal motion only when the ion motion is out-of-phase, the denominator in Eq.~(\ref{Eq265}) is proportional to $\omega^2_{o}$, instead of 
$c^2_l-c^2_t$. The role of the phonon force constants is to couple in-phase and out-of-phase motion at finite wavevectors.

Choosing Cu and O as typical cation and anion masses, we take the optical phonon frequency  of a $\text{CuO}_{2}$ plane, \textit{i.e.}, $\omega_{0} \sim 2.5$ THz~[\onlinecite{phononspectrum}], and  set   $\gamma_{1} \sim \gamma /2$ and $\gamma_{2}\sim 3\gamma /4$.   We find that the typical numerical value of $\eta_H$ at $B=10 \; \text{T}$  
is $\sim 5 \times  10^{-18} \; {\rm kg} \; {\rm s}^{-1}$.


The wave equations~(\ref{eqmotioneff1}) and~(\ref{eqmotioneff2}) explicitly couple longitudinal and 
transverse phonon modes. We characterize the chirality of acoustic phonon modes by the polarization $p_{a}$ 
defined as the ratio of the transverse to the longitudinal component in the dominantly longitudinal mode.
For wavevectors in the $\hat{x}$ direction $p_{a}\equiv|u_{ay}/u_{ax}|$. 
From Eqs.~(\ref{eqmotioneff1}) and~(\ref{eqmotioneff2}), 
\begin{align}
p_{a}\approx  \frac{c_{l} (k_{x}a)}{2\gamma} |\eta_{H}|\,,
\label{pol}
\end{align}
at long wavelengths. 
As shown by the overlapping blue lines in Fig.~\ref{Fig:2},  Eq.~(\ref{pol}) is in excellent agreement with the polarization obtained numerically from Eq.~(\ref{96newbasis}).  However, according to Eq.~(\ref{pol}), $p_{a}$ is non-zero only if $\eta_H \ne0 $, while  the polarization $p_{a}$ calculated from Eq.~(\ref{96newbasis}) vanishes at all wavevectors only when the elastic coupling between the in-phase motion and out-of-phase dynamics vanish entirely, \textit{i.e.} when $m_{+}=m_{-} $ and $\gamma_{1}=\gamma_{2}$.  The more stringent condition  
for the absence of field-induced coupling between longitudinal and transverse modes is captured when terms of order $(k_{x}a)^{4}B$ are retained in the low-energy effective model.

Figure~\ref{Fig:2} additionally displays the dependence of the polarization $p_{o}=|u_{oy}/u_{ox}|$ of the out-of-phase longitudinal mode on the magnetic field (orange line). Since  Lorentz forces directly couple  transverse and longitudinal out-of-phase modes, the polarization of the out-of-phase mode is several order of magnitude larger ($\sim 10^4$ for our parameters)  than the polarization of the in-phase mode. When the Lorentz force coupling the transverse and longitudinal out-of-phase motion vanishes, the polarization $p_{0}$ is significantly reduced (yellow line). We find that, however, $p_{0}$ does not vanish as long as there is a finite elastic coupling between in-phase and out-of-phase modes ($\gamma_{1} \neq \gamma_{2}$).  

In addition to coupling longitudinal and transverse  phonons, Lorentz forces also give rise to 
small changes in acoustic phonon frequencies. 
To leading order in the parameters $k_{x}a$ and  $\omega_{c}/\omega_{o}$, 
Eqs.~(\ref{eqmotioneff1}) and~(\ref{eqmotioneff2}) imply that both the longitudinal 
and transverse phonon frequencies are reduced by a factor of  
$1 - \omega_{c}^2/2\omega_{o}^2$. With our parameters, one finds $ \omega_{c}^2/\omega_{o}^2    \sim 10^{-9}$. 


\textit{Discussion and conclusions.} In this work, we have derived a low-energy effective model for the lattice vibrations of a ionic crystal subjected to a static magnetic field which accounts for the influence of Lorentz forces.
We find that the long-wavelength in-phase lattice dynamics is characterized by a finite phonon Hall viscosity, which implies chiral phonon transport. The Lorentz force contribution to the 
Hall viscosity is rooted in the coupling between the in-phase motion and out-of-phase motion 
of cations and anions at finite wavevectors.  This mechanism leads to typical values of the 
phonon Hall viscosity $\eta_{H} \sim 5 \times 10^{-18}$ kg s$^{-1}$ at magnetic field $B=10$ T.
In comparison Barkeshli \textit{et al.} investigated phonon coupling to a variety of time-reversal 
symmetry broken electronic states, and estimated [\onlinecite{1PhononHall}] resulting Hall viscosities in the range $\eta_{H} \sim 10^{-19} - 10^{-15} \; \text{kg} \; \text{s}^{-1}$.  

Our results apply to layered $3d$ crystals as well when considering phonons with wave-vector oriented parallel to the $2d$ layers. In the limit of vanishing interlayer coupling, one could obtain the $3d$ phonon Hall viscosity simply by  redefining $\eta_{H} \rightarrow \eta_{H}/a$.

Our estimates are based on a diatomic square lattice for analytical tractability; however, our conclusions are not particular to the system analyzed. We anticipate any ionic crystal subjected to an out-of-plane magnetic field to be characterized by a  finite phonon Hall viscosity.

The emergence of the phonon Hall effect in ionic crystals subjected to a static magnetic field has been investigated within a quantum treatment by Agarwalla \textit{et al.}~[\onlinecite{PhononHallquantum}].  Our classical approach yields an analytical expression for the Hall viscosity, is more transparent and provides better
insight into the parameters that control its strength, allowing comparisons with other 
phonon Hall viscosity mechanisms. 

Experimentally probing the phonon Hall viscosity has proven to be a challenging task. 
The renormalization of  the long-wavelength acoustic phonon spectrum is far below the resolution limits of conventional spectroscopic probes.  Moreover, such measurement would not carry any information 
about the sign of the Hall viscosity. 
However, it has been recently shown that a finite phonon Hall viscosity is responsible for  circular bifringence of transverse acoustic waves~[\onlinecite{1PhononHall2}].  Thus, it yields an acoustic Faraday rotation that can be probed via  acoustic cavity interferometry~[\onlinecite{faraday2},\onlinecite{faraday1}]. Another candidate probe is time-dependent x-ray diffraction, which allows to directly image acoustic phonon modes~[\onlinecite{xray}]. 

Naturally, the phonon Hall viscosity yields chiral phonon transport. 
In an ionic crystal with no magnetic order or symmetry-broken electronic states one might expect our reported phonon Hall viscosity to be the source of phonon Hall signals.  Recently, though,  it was shown that scattering on charged defects, which are common in ionic crystals, can also lead to skew scattering~\cite{extrinsic}. However, the resulting phonon Hall effect is predicted to display a temperature dependence $\propto T^{-1}$, 
in contrast to the $\propto T^{5}$ dependence of the signal ascribed to impurity scattering of acoustic phonons with  finite Hall viscosity~\cite{sachdev1}, which should make the two signals easily distinguishable.

\textit{Acknowledgements.}
The authors thank B. Ramshaw, L. Taillefer, M.-E. Boulanger, G. Grissonansche, S. Kivelson,  K. Behnia and H. Guo for helpful discussions and experimental motivation for this work. This work was supported by the U.S. Department of Energy, Office of Science, Basic Energy Sciences, under Award  DE-SC0022106, and by the National Science Foundation under Grant No. NSF DMR-2144086.

\end{document}